\documentclass[acmtog]{acmart}

\AtBeginDocument{%
  \providecommand\BibTeX{{%
    \normalfont B\kern-0.5em{\scshape i\kern-0.25em b}\kern-0.8em\TeX}}}

\setcopyright{acmlicensed}

\copyrightyear{2024}
\acmYear{2024}
\setcopyright{acmlicensed}\acmConference[SA Conference Papers '24]{SIGGRAPH Asia 2024 Conference Papers}{December 3--6, 2024}{Tokyo, Japan}
\acmBooktitle{SIGGRAPH Asia 2024 Conference Papers (SA Conference Papers '24), December 3--6, 2024, Tokyo, Japan}
\acmDOI{10.1145/3680528.3687619}
\acmISBN{979-8-4007-1131-2/24/12}

\setcitestyle{square}

\newcommand{\HL}[1]{{#1}}

\newcommand{\DEL}[1]{}
\newcommand{\HHL}[1]{{#1}}
\newcommand{\HHHL}[1]{{#1}}

\newcommand{\Cmt}[1]{\textcolor{blue}{\small{ // {#1} } }}

\usepackage{algorithm}
\usepackage{algorithmic}

\begin{document}

\title{gDist: Efficient Distance Computation between 3D Meshes on GPU}

\author{Peng Fan}
\affiliation{%
  \institution{Zhejiang University}
  \country{China}}
\email{fanpeng0103@zju.edu.cn}

\author{Wei Wang}
\affiliation{%
  \institution{Zhejiang University}
  \country{China}}
\email{wangweiydyk@zju.edu.cn}

\author{Ruofeng Tong}
\affiliation{%
  \institution{Zhejiang University}
  \country{China}}
\email{trf@zju.edu.cn}

\author{Hailong Li}
\affiliation{%
  \institution{Shenzhen Poisson Software Co., Ltd.}
  \country{China}}
\email{lihailong@poissonsoft.com}

\author{Min Tang}
\authornote{Corresponding author, \url{https://min-tang.github.io/home/gDist/}.}
\affiliation{%
  \institution{Zhejiang University, Zhejiang Sci-Tech University}
  \country{China}}
\email{tang_m@zju.edu.cn}

\renewcommand{\shortauthors}{P. Fan et al.}

\begin{abstract}
Computing maximum/minimum distances between 3D meshes is crucial for various applications, i.e., robotics, CAD, VR/AR, etc. In this work, we introduce a highly parallel algorithm (gDist) optimized for Graphics Processing Units (GPUs), which is capable of computing the distance between two meshes with over $15$ million triangles in less than $0.4$ milliseconds (Fig.~\ref{fig:rings}). By testing on benchmarks with varying characteristics, 
\DEL{the algorithm achieves one to three orders of magnitude speedups over prior CPU-based algorithms}
\HHL{the algorithm achieves remarkable speedups over prior CPU-based and GPU-based algorithms}
on a commodity GPU (NVIDIA GeForce RTX 4090). Notably, the algorithm consistently maintains high-speed performance, even in challenging scenarios that pose difficulties for prior algorithms.
\end{abstract}



\begin{teaserfigure}
\centering
  \includegraphics[width=0.7\textwidth]
  {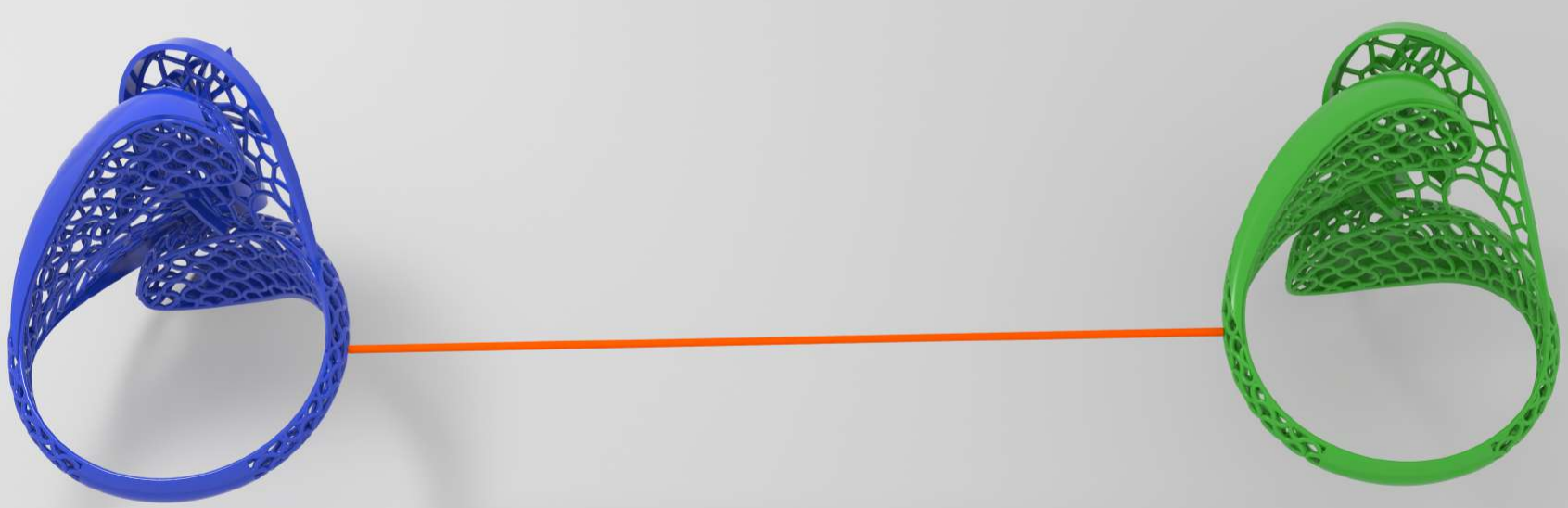}
  \caption{{\bf Benchmark Rings:} For the pair of rings, each consisting of $7.5M$ triangles, our algorithm can calculate the minimum distance between them (as indicated by the orange line) within $0.38$ milliseconds on an NVIDIA GeForce 4090. As compared to an optimized CPU implementation (PQP~\cite{Larsen14}), we observe $54X$ speedups for this specific frame and an average speedup of up to $26X$ across a sequence of rotation frames.}
  \label{fig:rings}
\end{teaserfigure}


\maketitle

\section{Introduction}
\label{sec:intro}

The precise computation of maximum/minimum distance between 3D models is crucial for a spectrum of applications, including robotics, computer-aided design (CAD), virtual reality (VR), and augmented reality (AR). Within CAD systems, distance computation is a key component for tasks such as interference checking and collision detection~\cite{Larsen14,Shellshear14}. By computing distances between models, CAD systems can rapidly discern potential conflicts, thereby amplifying the precision and dependability of designs. The practical ramifications of this technology extend broadly across diverse domains, encompassing manufacturing, engineering, architecture, physics simulation, and virtual reality. The perennial challenge lies in the optimization of distance computation efficiency, a conundrum persistently grappled with by both industrial and academic communities.

In recent years, Graphics Processing Units (GPUs) have emerged as formidable many-core parallel processors, extensively applied to accelerate various computationally intensive tasks, including surface tessellation~\cite{Xiong23}, mass property computation~\cite{Krishnamurthy08}, collision detection~\cite{CStreams11}, ray tracing~\cite{Meister20}, and physically-based simulation~\cite{pcloth20,Wang20}. However, notwithstanding these strides, the development of efficient GPU algorithms specifically tailored for distance computation remains \DEL{an unmet challenge} \HHHL{an challenge}.

\subsection{Main Results}

This paper addresses the efficiency challenge of computing maximum/minimum distances between 3D mesh models. We introduce a meticulously designed parallel algorithm optimized for GPUs, showcasing both efficiency and accuracy. Our algorithm demonstrates the capability to compute the distance between two meshes, totally consisting of over $15$ million triangles in less than $0.4$ milliseconds. Moreover, it achieves \DEL{a speedup of approximately one to three orders of magnitude faster than prior CPU algorithms} \HHL{remarkable speedups over prior CPU-based and GPU-based algorithms} on a commodity GPU (NVIDIA GeForce RTX 4090). The principal contributions of our work include:
\begin{itemize}
\item{{\bf Fine-grained parallel execution: } Our algorithm employs a fine-grained parallel design, achieving intricate load balancing and fully exploiting the parallel computational potential of GPUs for optimal performance.}
\item{{\bf Effective AABB-based culling:}  Through a novel AABB distance bounding formula, our algorithm introduces a more effective culling method, significantly reducing the computation load and enhancing overall performance.}
\item{{\bf Maximum and minimum distance computing: } Leveraging the unified culling algorithm and task decomposition strategy, our algorithm adeptly handles both maximum and minimum distance computing between 3D mesh models.}
\end{itemize}

These algorithms have been implemented on various GPUs and rigorously evaluated on complex benchmarks. As compared to prior approaches, our algorithm presents the following advantages:
\begin{itemize}
\item{{\bf Generality:} Our approach exhibits insensitivity to underlying motions, relative size, or distance between models. It accommodates both rigid and deformable models and computes both maximum and minimum distances.}
\item{{\bf Faster performance:} As compared to prior algorithms~\cite{Larsen14,Shellshear14,Pan12},  our approach demonstrates approximately one to three orders of magnitude speedups.}
\item{{\bf Lower memory overhead:} For scenes with $15$M triangles, the memory overhead is less than $980$ Mbytes, allowing our method to run efficiently on GPUs with limited capabilities.}
\end{itemize}

\section{Related Work}
\label{sec:related}

There are many efficient algorithms~\cite{Gilbert88,Lin91,Cameron97,Gino99} for determining the closest points between two convex objects. Quinlan~\shortcite{Quinlan94} proposed the utilization of a hierarchy of convex bounding volumes for the representation of general non-convex objects. A comprehensive survey on collision detection and distance computation is presented in ~\cite{Lin17}.

Larsen et al.~\shortcite{Larsen1999FastPQ} introduced a general paradigm for computing the minimum distance between non-convex objects represented in the bounding volume hierarchies (BVH) of sphere-swept bounding volumes. Their method maintains a priority queue of pairs of BVH nodes, where each BVH node serves as the root for a subtree, representing a volume bounding a subset of the object. Johnson et al.~\shortcite{Johnson1998AFF,Johnson2005MinimumDQ} adopted a similar approach to compute the minimum distance between two B-spline surfaces. Chang et al.~\shortcite{CHANG11} and Kim et al.~\shortcite{Kim11} proposed algorithms for distance computation between B\'ezier surfaces and B-spline surfaces, respectively. Son et al.~\shortcite{Son20} empolyed toroidal patches for minimum distance computation for solids of revolution.

The proximity query package (PQP)~\cite{Larsen1999FastPQ} is considered one of the most efficient algorithms for distance query between 3D mesh models, utilizing rectangular swept sphere (RSS) trees as the acceleration data structure. 
FCPW~\cite{Sawhney23} and Embree~\cite{Embree16} represent attempts at further acceleration, \HL{demonstrating key performance enhancements concerning point-to-mesh distance queries.}
Shellshear and Ytterlid~\shortcite{Shellshear14} employed SSE instructions for acceleration on CPUs, resulting in about a $10\%$ reduction in overall query time. Pan et al.~\shortcite{Pan12} released an open-source package for collision detection and distance queries, primarily used in robot simulation.

Most algorithms for distance queries between 3D models are based on bounding volume hierarchies (BVHs)~\cite{Lauterbach10,Larsen1999FastPQ,Shellshear14}. As models undergo deformation, these hierarchies are either updated or reconstructed. In order to accelerate the computations on BVH, several parallel algorithms utilizing multiple cores on a CPU or GPU have been proposed~\cite{PKS10,Kim09,Heo10,TMT10,FWPS11,Zhang14,WDZ17}. Many BVH-based parallel algorithms maintain or update a BVTT (Bounding Volume Traversal Tree) front to accelerate computations~\cite{Klosowski98,Li98,TMT10,CStreams11,Zhang14,BVHCD18,Chitalu20}. These parallel algorithms maintain a BVTT front for parallel collision checking, which can incur a substantial memory overhead.

Recently, Zong et al.~\shortcite{Zong23} introduced an efficient algorithm for point-to-mesh distance queries. However, extending this approach to distance queries between two meshes proves to be challenging.

\section{Algorithm Pipeline}
\label{sec:algorithm}

Mathematically, given two 3D models, $P$ and $Q$ in $\mathbb{R}^3$, the maximum and minimum distances between them can be defined as~\cite{Lin17}:
\begin{eqnarray}
    D_{min} &=& \mathrm{min}_{p \in P} \mathrm{min}_{q \in Q} 
    \quad \mathrm{Dist}(p, q),\\
    D_{max} &=& \mathrm{max}_{p \in P} \mathrm{max}_{q \in Q} 
    \quad \mathrm{Dist}(p, q),
\end{eqnarray}
where $\mathrm{Dist}$ is the distance function, which can be any distance norm. Most standard applications employ the Euclidean distance norm. $D_{min}$ and $D_{max}$ denote the separation distance and the spanning distance between the two models, respectively.

\HL{We construct Axis-Aligned Bounding Box (AABB) hierarchies~\cite{Bergen97} for the given models, utilizing them as acceleration data structures. In Section~\ref{sec:bvh}, \DEL{we introduce an optimized BVH tailored for improved GPU performance} \HHL{we design a data structure called f12-BVH \HHHL{(a full binary tree with 1 or 2 triangles on each leaf node)} which is an optimized BVH tailored \HHHL{to improve} GPU performance}. The f is from "full", 1 and 2 are from the "1 or 2 triangles. Additionally, we maintain BVTT fronts for fine-grained task decomposition and a reduction in the number of
pairwise bounding volume (BV) tests~\cite{CStreams11,BVHCD18,Chitalu20}.}

The algorithm's overarching structure is delineated in Algorithm~\ref{alg:overall}. The procedure begins by initializing two buffers ($\rm buffer_1$ and $\rm buffer_2$), which represent the BVTT fronts, along with an initial estimate for the upper and lower bounds of the minimum distance between two AABBs (lines 1-2, Section~\ref{sec:AABB}). 

$\rm buffer_1$ exclusively contains the BVTT front composed of the two root nodes of the BVHs, and the estimate $\rm upper_{global}$ is set as the upper bound of the minimum distance between the BVs of the two root nodes (lines 3-4). The algorithm then iteratively expands the BVTT in the buffer until all BVH nodes within the BVTT are leaf nodes (lines 5-9, Section~\ref{sec:bvtt}), ultimately returning $\rm upper_{global}$ as the resultant $D_{min}$ (lines 10-11).

\begin{algorithm}[t]
\caption{Parallel Minimum Distance Computing} 
	\label{alg:overall} 
	\begin{algorithmic}[1]
        \renewcommand{\algorithmicrequire}{\textbf{Input:}}
		\REQUIRE {Two BVHs: $root_{A}, root_{B}$}
	  \renewcommand{\algorithmicensure}{\textbf{Output:}}
		\ENSURE {The minimum disance: $D_{min}$}
            \STATE Initialize $buffer_{1}, buffer_{2}$ \Cmt{BVTT fronts}
            \STATE $lower, upper \gets$ calculateDistanceBounds$(root_A, root_B)$
            \STATE $buffer_{1} \gets root_{A},root_{B},lower,upper$ \Cmt{Update the front}
            \STATE $upper_{global} \gets upper$
            \WHILE{Not Reach Leafs}
                \STATE $step \gets$ calculateAdaptiveDepth($buffer^{length}_{1}$)
                \STATE $buffer_{2} \gets$ expandBvtt$(buffer_{1}, upper_{global}, step)$
                \STATE swapBuffer($buffer_{1},buffer_{2}$)
            \ENDWHILE
            \STATE $D_{min} \gets upper_{global}$
            \RETURN $D_{min}$
\end{algorithmic} 
\end{algorithm}

In each iteration, an adaptive expansion depth step is computed based on the current buffer size (line 6, Section~\ref{sec:depth}). Subsequently, all BVTT nodes in the existing buffer are expanded, producing a set of new BVTT nodes stored in another buffer (line 7). The two BVTT buffers are then switched using a double-buffer mechanism (line 8).

Symmetrically, the algorithm can be modified to calculate the maximum distances between 3D mesh models,
as detailed in the supplementary material. 
\HL{Sections~\ref{sec:AABB} through to \ref{sec:depth} will be dedicated exclusively to minimum distance computation. Similar methodologies can be derived for the maximum distance computation.}

\section{Enhanced Distance Bounds for AABBs}
\label{sec:AABB}

A key component of our algorithm is the utilization of AABB distance bounds for culling, as demonstrated in Algorithm~\ref{alg:overall}. Here, we derive enhanced distance bounds for AABBs, which are \HL{superior} to conventional bounds.

For a given AABB pair, i.e., \HL{$V_a$ and $V_b$, where:
\begin{equation}
\begin{split}
    V_a = &\{P | A^{min}_i\leq P_i \leq A^{max}_i, i=1,2,3\}
\end{split}
\end{equation}
\begin{equation}
\begin{split}
    V_b = &\{P | B^{min}_i\leq P_i \leq B^{max}_i, i=1,2,3\}
\end{split}
\end{equation}
}

\begin{figure}[]
\centering
\includegraphics[width=0.85\linewidth]{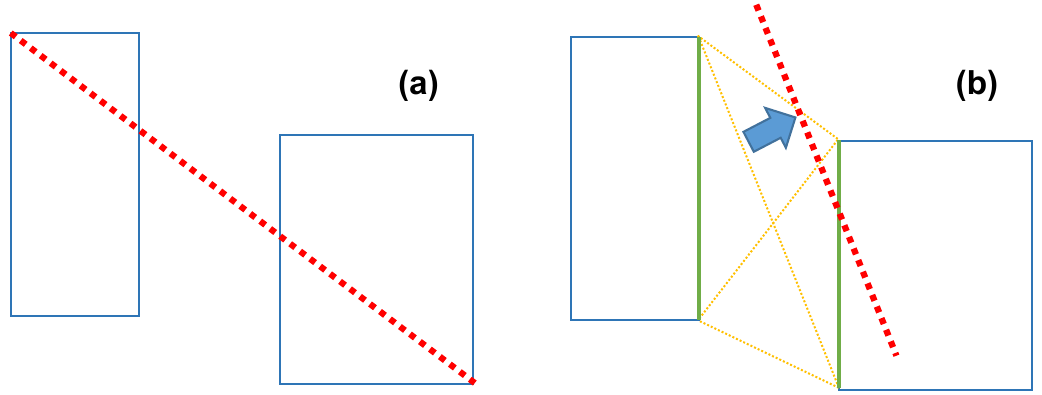}
\caption{{\bf Distance Bounds for AABBs:} Compared to the conventional bound computation (a), our new method returns a tighter bound (b). The upper bounds for the minimum distance are highlighted in red.}
\label{fig:bound}
\end{figure}

The conventional bounds can be defined as follows
\DEL{[Congusbongus 2018]}
\HHL{~\cite{Krishnamurthy11}}. 
The lower bound of the distance between them, $d_{min}^{L}$ can be defined as:
\begin{equation}
\begin{split}
d_{min}^{L} = \sqrt{\sum_{i=1,2,3}f_{min}(A^{min}_i, A^{max}_i,B^{min}_i, B^{max}_i)^2},
\end{split}
\end{equation}
where function $f$ returns the projected distance between two AABBs on a specific axis, i.e.:
\begin{equation}
\begin{split}
&f_{min}(A^{min}_i, A^{max}_i,B^{min}_i, B^{max}_i)=\\&\begin{cases}
0 &\HL{\text{if the projections overlap}}\\
min\{|A^{min}_i-B^{max}_i|, |A^{max}_i-B^{min}_i|\}  & \HL{\text{otherwise}}
\end{cases}
\end{split}
\end{equation}

Symmetrically, the upper bound of the distance between the two AABBs can be defined as:
\begin{equation}
\label{eq:dum}
\begin{split}
d_{max}^{U} = \sqrt{\sum_{i=1,2,3}f_{max}(A^{min}_i, A^{max}_i,B^{min}_i, B^{max}_i)^2},
\end{split}
\end{equation}
where:
\begin{equation}
\begin{split}
&f_{max}(A^{min}_i, A^{max}_i,B^{min}_i, B^{max}_i)=\\
&max\{|A^{min}_i-B^{max}_i|, |A^{max}_i-B^{min}_i|\} 
\end{split}
\end{equation}

\HL{A conventional upper bound of the minimum distance between AABBs, $\hat{d}_{min}^{U}$, is determined by the fastest points contained by them, which is equal to the upper bound of the maximum distance:
\begin{equation}
\begin{split}
\hat{d}_{min}^{U} = max_{P\in V_a, Q \in V_b} ||PQ|| = d_{max}^{U}
\end{split}
\end{equation}
}

Although these conventional bounds are widely used, we see potential for further enhancements to increase culling efficiency. We define an AABB as `tight' only if a vertex of its encompassing model aligns with a point on each of its six bounding rectangles.

If $V_a$ and $V_b$ are tight, we obtain the following enhanced bounds:
\HL{
\begin{equation}
d_{min}^{U} = min\{max_{P\in S^{V_a}_i, Q \in S^{V_b}_j} ||PQ||\},
\end{equation}
\begin{equation}
d_{max}^{L} = max\{min_{P\in S^{V_a}_i, Q \in S^{V_b}_j} ||PQ|| \}, 
\end{equation}
where $S$ represents the six bounding rectangles of each AABB, and $i=1,2,...,6, j=1,2,...6$. }\HL{A comparison between $\hat{d}_{min}^{U}$ (left) and $d_{min}^{U}$ (right) is demonstrated in Fig.~\ref{fig:bound}.}

It can be proven that during the maintenance of a BVH, if all its leaf nodes' AABBs are tight, then all the nodes' AABBs are tight.
The overestimation of the upper bound for the minimum distance and the underestimation of the lower bound for the maximum distance do not affect the correctness of the algorithm. Hence, it is not necessary to calculate all $36$ pairs of rectangular comparisons. However, in practice, due to the well-hidden cost during the computation process, we choose to compute all $36$ pairs of rectangular comparisons.

It is noteworthy that in some applications, it is a common practice to slightly enlarge the bounding box to improve the robustness of the algorithm. However, in this case, employing a similar approach would loosen the bounding box, resulting in underestimated $d_{min}^{U}$ or overestimated $d_{max}^{L}$, leading to incorrect culling.

\section{BVTT Expansion}
\label{sec:bvtt}

\begin{algorithm}[t]
\caption{Expand BVTT Buffer} 
	\label{alg:BVTT} 
	\begin{algorithmic}[1]
        \renewcommand{\algorithmicrequire}{\textbf{Input:}}
	    \renewcommand{\algorithmicensure}{\textbf{Output:}}
		\REQUIRE $buffer_{input}, upper_{global}, step$
		\ENSURE $buffer_{output}$
            \STATE Initialize $buffer_{output}$
		\FOR{each $BVTT \in buffer_{input}$}
                \STATE $children_{A} \gets$ getChildren$(BVTT_A, step)$
                \STATE $children_{B} \gets$ getChildren$(BVTT_B, step)$
                \FOR{each $child_A \in children_{A}$}
                    \FOR{each $child_B \in children_{B}$}
                        \STATE $lower, upper \gets$ calculateDistanceBounds$(child_A, child_B)$
                        \IF{$lower < upper_{global}$}
                            \STATE $bufferd_{output} \gets child_A,child_B,lower$
                            \STATE $upper_{global} \gets$ min$(upper_{global},upper)$
                        \ENDIF
                    \ENDFOR
                \ENDFOR
            \ENDFOR
            \RETURN $buffer_{output}$
	\end{algorithmic} 
\end{algorithm}

Each BVTT node encapsulates two nodes from the BVH trees of two objects, signifying that this BVTT is designed to compute the distance between the objects contained in these two nodes. Initially, there exists a single BVTT node comprising the root nodes of two BVH trees. Typically, a BVTT node can be expanded into four new BVTT nodes, with these four nodes originating from permutations and combinations of the left and right subtrees of the nodes from the two BVH trees.

\begin{figure}[t]
	\centering
	\includegraphics[width=0.8\linewidth]{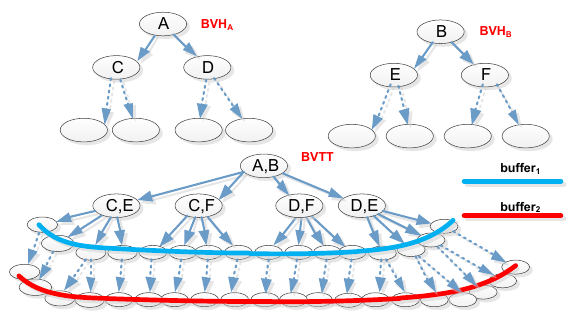}
    \caption{{\bf BVTT expansion:} We traverse all BVTT nodes in the input front $\rm buffer_1$ and, based on the input adaptive depth step, obtain the expanded BVTT front $\rm buffer_2$.}
	\label{fig:bvtt}
\end{figure}

The procedure for expanding the BVTT in the buffer is outlined in Algorithm~\ref{alg:BVTT}. It traverses all BVTT nodes in the input buffer and, guided by the adaptive depth step (explained in Section~\ref{sec:depth}), generates descendants for the two BVH nodes contained in the BVTT (lines 3-4). Subsequently, it iterates through all combinations of these descendants (lines 5-13). For each pair, it computes the upper and lower bounds of the minimum distance between the bounding boxes. If the current lower bound of the minimum distance for this pair of bounding boxes surpasses the ongoing (global) estimate of the upper bound of the minimum distance, it implies that this pair of bounding boxes can be disregarded without further expansion. Otherwise, it needs to be appended to the output BVTT buffer (line 9), and the upper bound of the minimum distance for this pair of bounding boxes is used to update the global estimate of the upper bound of the minimum distance (line 10). An illustrative example of BVTT expansion is presented in Fig.~\ref{fig:bvtt}.

\section{Adaptive Expansion Depth Computation}
\label{sec:depth}

In the context of distance computation, sequential algorithms typically employ a heuristic depth-first search (DFS) strategy that swiftly identifies closely located primitives, potentially terminating upon reaching a leaf node. To illustrate, consider an extreme case where the sequential algorithm discovers the optimal solution upon first reaching a leaf node,  enabling immediate termination. In contrast, parallel algorithms based on Breadth-First Search (BFS) typically iterate one layer at a time, with each thread corresponding to a node. When the sequential algorithm rapidly finds the correct answer through heuristic depth-first search, it implies effective pruning for parallel algorithms. In such scenarios, the buffer contains only a small number of BVTT nodes, leading to underutilization of GPU hardware resources, as each iteration involves launching only a limited number of threads to process these BVTT nodes. Moreover, the traditional strategy's extensive unfolding requires frequent CPU setting adjustments based on the GPU's previous computation results, resulting in substantial CPU-GPU synchronization and a consequent drop in algorithm performance.

A direct solution is to expand multiple layers at once. In~\cite{Lauterbach10}, the author experimented with converting BVH into an octree, essentially expanding three layers in BVH. However, adopting a fixed number of expansion layers is suboptimal, as it may lead to either too few or too many BVTT nodes during unfolding. If there are too few nodes, the aforementioned issues persist. Conversely, if there are too many nodes, expanding multiple layers at once might compromise pruning efficiency. A BVTT node, which could have been pruned after the first layer, might be expanded to the third layer, resulting in wasted space and time.

In essence, a dynamic adjustment of the number of layers expanded is required based on the number of nodes. When the buffer contains $n$ BVTT nodes, the adaptive unfolding algorithm aims for a constant $C$. Specifically, it seeks the maximum expansion depth $k$ such that: 
$2^{2k}n < C.$
\HHHL{Since our f12-BVH is a full binary tree, expanding the $n$ BVTT nodes by $k$ levels will result in fewer than $C$ BVTT nodes.
With the depth $k$, the BVTT nodes are expanded in parallel on the GPU. 
In practice, $C$ is typically set to 1024×256 for optimal performance, and $k$ is chosen to be 5, corresponding to five expansion levels. }

\section{Parallel Distance Computation on GPU}
\label{sec:parallel}

\subsection{Parallel Culling}

In scenarios where the quantity of BVTT nodes grows exponentially in the outlined algorithm, the challenge arises in controlling their number by culling some of them. This culling process hinges on estimating the maximum value for the minimum distance between bounding boxes.

In each iteration, we calculate the minimum maximum distance between the bounding boxes of the two BVH nodes in every BVTT node obtained. Subsequently, we execute a reduction operation to identify the minimum among these maximum values. This minimum value is then used as the estimate for the minimum distance. All BVTT nodes with bounding box distances exceeding this estimated distance can be efficiently eliminated in the subsequent iteration. This culling mechanism ensures a controlled and optimized progression of BVTT nodes throughout the algorithm.

\subsection{Throughput Consideration}

In the context of our parallel algorithm, a pertinent question arises: How can we harness the full parallel high-throughput advantages of the GPU when the number of BVTT nodes is low?

Traditional traversal methods often assign one thread per existing BVTT node, with each thread responsible for the expansion of a single BVTT node. While this approach is standard, especially in distance calculations with the elimination algorithm described earlier, it tends to underutilize the GPU's resources as only a limited number of threads are processed in each kernel call.

Our algorithm introduces a novel strategy by allocating one BVTT node to multiple threads. In this design, each thread corresponds to a newly generated node after expanding the associated BVTT node. For instance, when expanding one layer, if both BVH nodes have left and right children, four possible combinations arise. Consequently, four threads process one BVTT node, with each thread potentially generating a new BVTT. While expanding only one layer is insufficient for optimal GPU resource utilization, the algorithm dynamically determines the number of layers to expand based on the current BVTT node count. For a given number n of BVTT nodes and expanding $k$ layers, the maximum number of newly generated nodes is at most $2^{2k}n$. The algorithm dynamically adjusts $k$ to ensure the generated node quantity falls within an appropriate range. \HHHL{Please refer to Section~\ref{sec:depth} for the calculation of parameter $k$.}

This parallel approach enables the algorithm to maximize GPU resource utilization, launching the maximum number of threads each time a BVTT node is expanded. Particularly in simpler tasks, this strategy reduces the number of expansion iterations, enhancing overall algorithm efficiency.

\subsection{F12-BVH}
\label{sec:bvh}

To implement the described algorithm on the GPU, adjustments to the BVH structure are necessary. The commonly used lbvh~\cite{Karras12,Lauterbach09} proves suboptimal for such operations due to the non-trivial task of finding the $k-th$ descendant, requiring sequential searches based on left and right child pointers. This complicates determining which pair of descendants from two BVH nodes in a BVTT node should be processed by a given thread.

\begin{figure}[t]
\centering
\includegraphics[width=0.9\linewidth]{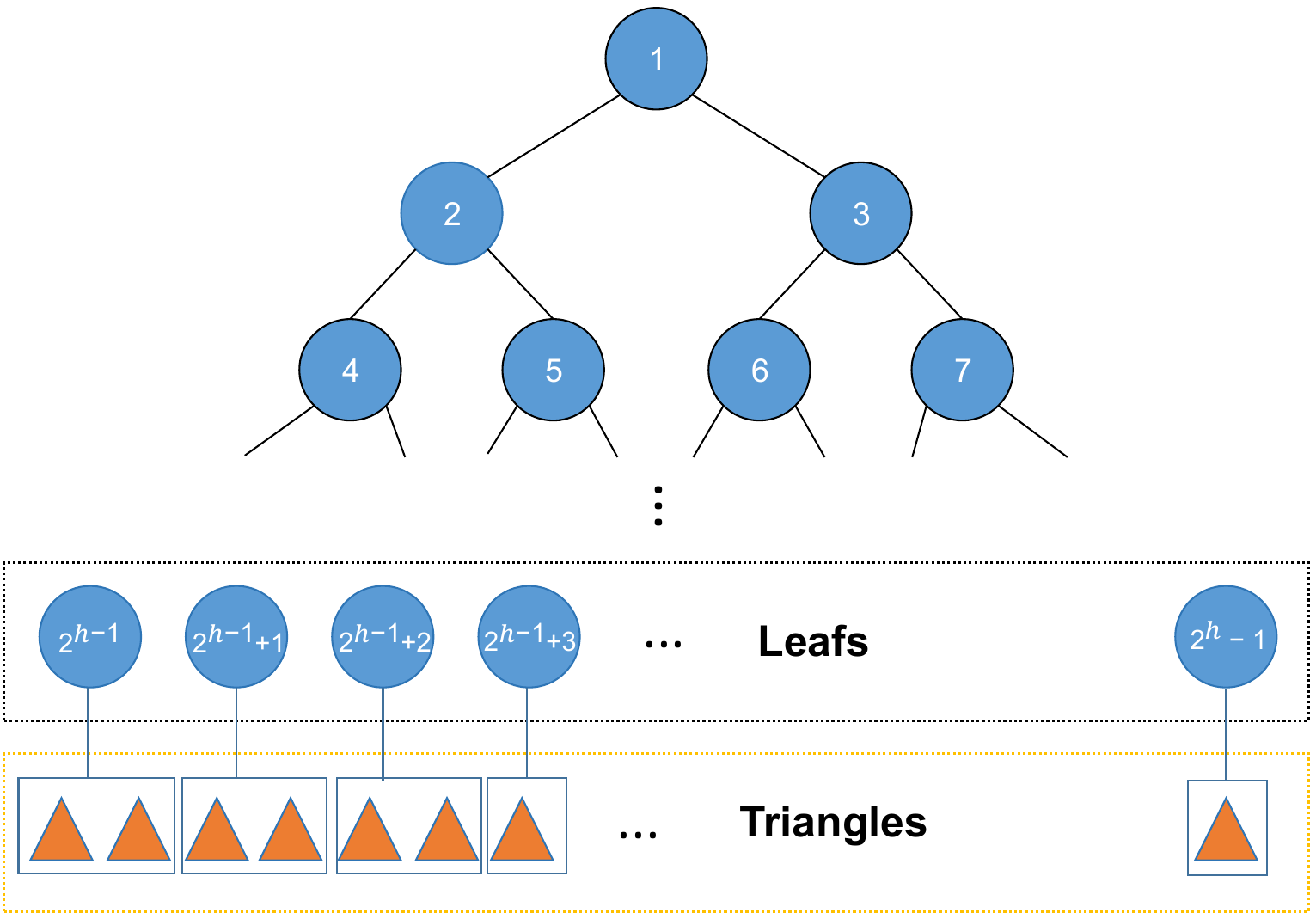}
\caption{{\bf F12-BVH:} We propose f12-BVH, a full binary tree with 1 or 2 triangles on each leaf node, designed for enhanced GPU performance.}
\label{fig:bvh}
\end{figure}

To address this, we propose f12-BVH, designed for enhanced GPU performance, as depicted in Fig.~\ref{fig:bvh}. Constructed on the BV sequence arranged by \HHL{Morton code}~\cite{Lauterbach09}, we treat adjacent primitives as one in cases where the number of BVs is not exactly a power of two, adjusting the number of primitives to a power of two. While this may slightly impact BVH quality, the advantages outweigh the trade-offs. 
The construction of the f12-BVH can be facilitated using a greedy algorithm, as detailed in the supplementary material.

\HHL{
Our algorithm does not rely on any assumption about the triangle order. While our algorithm may not perform as well on extremely poor spatial partitioning. We ensure high efficiency even under less favorable conditions, as demonstrated in benchmarks such as the truck, which has dramatic variations in triangle distribution and size, and the intersection, where the triangles of the two objects are closely packed together.
}

Accessing the $k-th$ descendant of BVH nodes in a full binary tree is straightforward using bitwise operations. This structure also allows us to determine, based on the thread ID, which newly generated BVTT node the current thread should handle corresponding to the BVTT expansion. In a scenario with $n$ BVTT nodes to be expanded to $k$ layers, the thread with ID $t$ should handle the $\lfloor \frac{t}{2^{2k}} \rfloor -th$ BVTT. The lower $2k$ bits of $t$ encode which descendants of the BVH nodes in BVTT will form a new BVTT node.

A full binary tree can be implicitly stored, and BVH nodes can contain only bounding box information. Our f12-BVH can completely omit any structure-related information, impacting BVH quality but offering advantages. In a full binary tree, no node lacks a left or right child, making each thread perform identical tasks. Threads obtain bounding boxes of two descendants based on the thread ID, calculate the distance between bounding boxes, and decide whether to output the bounding box. Specifically, each thread initially accesses the relevant BVTT. It's noteworthy that adjacent threads often access the same or neighboring BVTT positions, creating coalesced memory access on GPUs. Following this, the threads proceed to access the bounding boxes of the two descendants. In practical terms, the number of layers $k$ expanded at once is frequently substantial. Consequently, adjacent threads predominantly process neighboring nodes of BVH nodes contained within the same BVTT. This characteristic contributes to an efficient and parallelized execution on the GPU.
As the full binary tree is naturally stored in BFS order in memory, the algorithm proves cache-friendly.

\subsection{Mantainance of the Minimum Value}

Our pruning method hinges on estimating the upper bound of the minimum distance, where the minimum value is derived from the upper bounds calculated by all BVTT nodes.

We adopt a block-wise reduction strategy to maintain the minimum value: Threads store the calculated upper bounds in shared memory, perform a reduction operation within each block, and then one thread in the block uses atomic operations to update the global minimum distance upper bound.
It's important to note that because the CUDA-provided atomic operation \textbf{atomicMin} doesn't support floating-point types, the atomic operations mentioned above must be implemented manually \HL{via \textbf{atomicCAS} for integers.}

\section{Implementation and Results}
\label{sec:results}

We implemented our algorithm on three commodity GPUs, an NVIDIA RTX 2060 (featuring $1960$ CUDA cores at $1.7$GHz and $6$GB memory), an NVIDIA GeForce RTX 3090 (with $10496$ CUDA cores at $1.7$GHz and $24$GB memory), and an NVIDIA GeForce RTX 4090 (with $16384$ CUDA cores at $2.52$GHz and $24$GB memory).  These GPUs, characterized by varying core counts, were employed to assess the parallel performance of our approach. The implementation is developed using the CUDA toolkit version 11.7, and Visual Studio 2022 serves as the underlying development environment. All computations on the GPU are conducted using single-precision floating-point arithmetic. The testing environment involves a standard PC running Windows 11 Ultimate 64-bit, equipped with an Intel \HL{i5-12400f} CPU operating at $2.5$GHz and $16$GB RAM.

As part of preprocessing, we construct the AABB-based BVH (i.e., f12-BVH) on the CPU. Additionally, for scenes involving rotation motions or deformations, we perform BVH refitting to ensure the algorithm's adaptability to dynamic scenarios.
The resulting minimum distance in the presence of penetration is zero. For all our benchmarks, we adhered to a uniform distribution of time intervals.

Our performance evaluation uses 10 diverse benchmarks, each targeting a specific aspect of our algorithm's capabilities: 
\begin{itemize}
  \item {\bf Rings:} Two rings are rotating around their respective axes. Each ring, defined by an axis passing through its local center, comprises $7.5$M triangles (Fig.~\ref{fig:rings}).
  \item {\bf Apples:} Two Voronoi apples are moving apart from each other. Each apple model consists of $720$K triangles. (Fig.~\ref{fig:bench-3}).
  \item {\bf Tools:} This benchmark features a tool moving in parallel to another tool. Each tool comprises $1$M triangles (Fig.~\ref{fig:bench-1}).
  \item {\bf Comet-Tool:} This benchmark involves a comet model, comprising $1$M triangles, moving through a hole in a tools model, also with $1$M triangles (Fig.~\ref{fig:bench-5}). Importantly, the sizes of the two models significantly differ.
  \item {\bf Balls:}  A smaller Voronoi ball is enclosed by a larger one, and both are rotating synchronously about the same axis (Fig.~\ref{fig:bench-6}). Each ball model is made up of $130K$ triangles.
  \item {\bf Penetration:} A ratchet (with $1$M triangles) moves through a hole in a base tool (also with $1$M triangles)  (Fig.~\ref{fig:bench-7}). Notably, the two tools are at the same scale.
  \item \HL{{\bf Terrains:} Two terrain meshes (each with $9.7K$ triangles) are in close proximity to one other (Fig.~\ref{fig:bench-12}).

  }
  \item {\bf MaxDist:}  This benchmark mirrors the configuration of the Tools benchmark. However, in this case, the focus is on computing the maximum distance instead of the minimum distance (Fig.~\ref{fig:bench-8}).
  \item \HL{{\bf Truck:} A truck ($1.1$M triangles) is traveling between two rows of shelves  ($2.41$M triangles) (Fig.~\ref{fig:bench-11}).
  This benchmark consists of objects with non-uniform tessellation and poorly shaped triangles (large and slanted triangles on the shelves). 
  }
  \item {\bf Intersection:}  A challenging scene contains two triangle clusters. Each cluster comprises $1$ million triangles and intersects at a pre-specified point (Fig.~\ref{fig:bench-10}).
\end{itemize}

All the benchmarks are computing the minimum distance between models except the MaxDist benchmark.
The tool and the comet models are sourced from ~\cite{Krispel18}, while the Voronoi ball, apple, and ring are downloaded from \url{https://www.cgtrader.com/}.

\subsection{Performance}

\begin{figure}[t]
	\centering
 \includegraphics[width=0.88\linewidth]{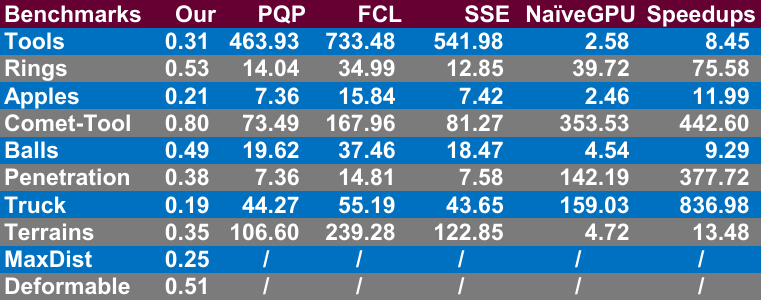}
    \caption{{\bf Performance Comparison:} we conduct a comprehensive comparison with several optimized CPU algorithms (PQP~\cite{Larsen14}, FCL~\cite{Pan12}, and SSE~\cite{Shellshear14}) \HHHL{and a `na\"ive' GPU implementation (based on ~\cite{BVHCD18})} \HL{(quantified in milliseconds)}. Across all benchmarks, our GPU algorithms consistently exhibit remarkable speedups, 
    notably outperforming the tested algorithms on an NVIDIA GeForce RTX 4090. 
	}
	\label{fig:performance-all}
\end{figure}

\HHHL{
A comparative analysis \HL{(Fig.~\ref{fig:performance-all})} was conducted against several optimized CPU algorithms, namely PQP~\cite{Larsen14}, FCL~\cite{Pan12}, and SSE~\cite{Shellshear14}, and a `na\"ive' GPU implementation derived from~\cite{BVHCD18}. 
}

\begin{figure}[t]
	\centering
	\includegraphics[width=0.85\linewidth]{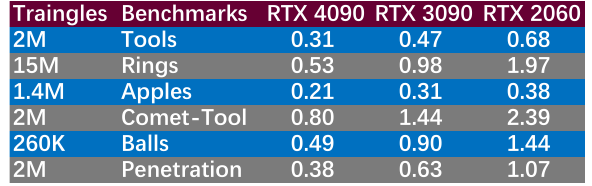}
    \caption{{\bf Performance on GPUs:} We rigorously evaluate the performance \HL{(quantified in milliseconds)} of our GPU-based algorithm on three commodity GPUs, establishing its efficacy across diverse GPU architectures.
	}
	\label{fig:performance-gpu}
\end{figure}

\HHHL{
All the CPU algorithms utilize rectangular swept sphere (RSS) trees but employ different elementary test approaches such as triangle-triangle distance computation, GJK algorithm~\cite{Gilbert88}, and SSE-accelerated algorithm. 
Since we haven't identified any publicly available GPU implementations for comparison, we developed a `na\"ive' GPU solution derived from~\cite{BVHCD18}. In this strategy, each thread on the GPU coordinates with a triangle in one model and employs a Depth-First Search (DFS)-like method for traversing the other model's BVH tree. In a CPU single-thread algorithm, it is straightforward to utilize the transient results generated in the DFS process for culling. However, the DFS-like algorithm on GPU necessitates each thread to use atomic operations for modifying/accessing a global variable that stores the current minimum/maximum distance.
}

In practice, we found that this DFS-like algorithm not only traverses more BVTT nodes (at least nine times, usually dozens of times, in all our benchmarks) but also easily causes severe thread load imbalance issues. We noticed that in some benchmarks, few threads searched too many leaf nodes, leading to a hundredfold decrease in the efficiency of the DFS algorithm. This is often caused by the characteristics of the scene itself. Therefore, although this GPU algorithm achieved improved performance in some scenarios, it could potentially be even slower than existing CPU methods, such as PQP, as depicted in Fig.~\ref{fig:performance-all}. 
Across all benchmarks, our GPU algorithms achieved noteworthy speedups over the `na\"ive' GPU solution ranging from $8.45-836.98X$.

Our algorithm achieved significant speedups in the Comet-Tools and Penetration benchmarks ($442.6$X and $377.72$X, respectively), thanks to the GPU's parallelization, which efficiently handles numerous triangle pairs with the same minimum distance.

\HHHL{
The Truck and Terrains benchmarks demonstrate that our algorithm achieves excellent performance for scenarios involving objects with non-uniform tessellation, or scenes containing meshes in a very close distance (such as in the calculation of Hausdorff distance), reporting about $836.98$X and $13.48$X speedups, respectively. These are common cases in engineering practices, which have verified the generality of our algorithm.
}

\HHHL{
The MaxDist benchmark, designed for maximum distance computing, is not supported by the CPU algorithms mentioned above. Our GPU algorithm showcased an average query time of approximately $0.25ms$ per frame in this scenario. Additionally, the Deformable benchmark underscores our algorithm's adaptability to scenes featuring deformable objects, achieved through dynamic refitting of our AABB-based BVH on-the-fly. In contrast, CPU algorithms face challenges due to prolonged waiting times for updating their underlying bounding volumes (RSSes), rendering them impractical for such dynamic scenes.
}

\HHHL{
Our algorithm undergoes testing under a worst-case scenario configuration, exemplified by Benchmark Intersection~~\cite{Krispel18}. In this particular scene, each model comprises $1$ million triangles and intersects at a pre-specified point. 
This configuration poses a formidable challenge for the `na\"ive' GPU solution, requiring approximately $3778$ seconds for minimum distance computation between the two models. This scenario is highly specialized, where the BVH itself plays a minimal role.
In contrast, our GPU algorithm accomplishes the same task in a mere $0.59$ milliseconds, showcasing a remarkable speedup of above $6,403$ times. This emphasizes the robustness and efficiency of our algorithm, especially in scenarios that prove highly challenging for conventional CPU-based and GPU-based approaches.
}

\HHHL{
Fig.~\ref{fig:performance-gpu} provides an overview of the triangle count across various benchmarks, emphasizing the efficacy of our algorithm in diverse scenarios. The average query time of our GPU-based algorithm on three distinct GPUs (NVIDIA RTX 4090, 3090, and 2060) underscores its compatibility with different GPU architectures, showcasing performance scalability in tandem with the number of CUDA cores.
}

\begin{figure}[t]
	\centering
	\includegraphics[width=0.85\linewidth]{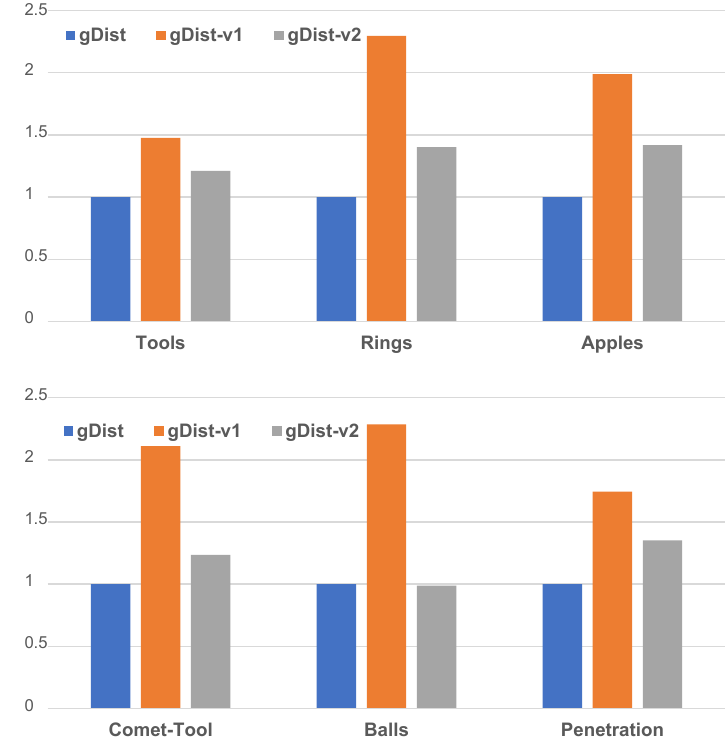}
    \caption{\HL{{\bf Ablation Study 1:} The figure presents an ablation study to highlight the contribution of each component in our method. 
    }}
	\label{fig:ablation}
\end{figure}

\HL{
\subsection{Ablation Studies}
Figure~\ref{fig:ablation} presents an ablation study to highlight the contribution of each component in our method. For each benchmark, we sampled three implementations for performance comparison: gDist is our proposed algorithm, gDist-v1 is a variant of gDist where the enhanced distance bound computation for AABBs is turned off, and only conventional bound calculation is utilized, and gDist-v2 is another variant of gDist where adaptive BVTT expansion is disabled. In the figure, we normalize the running time of gDist as $1$, and display the running times of gDist-v1 and gDist-v2 as ratios over gDist. It can be observed that in different benchmarks, disabling the enhanced bound calculations causes a running time increase of $1.5-2.2$X. Similarly, disabling the adaptive BVTT expansion based on f-12 BVH also results in performance deterioration in nearly all scenarios. 
These ablation studies have clearly illustrated the contributions of different components, and the final performance gain we have achieved is the combined result of these components.
}

Figure~\ref{fig:ablation2} presents another ablation study: Through the use of conventional distance bounds and our enhanced bounds, we can observe the changes in the number of BVTT nodes. We set the number of nodes when using the conventional bounds as $1$, and the number when using the enhanced bounds as its proportion. The culling rates in all scenarios are less than $45\%$. In Benchmark Apples, the number can be less than $4\%$. This experiment clearly demonstrates how the enhanced bounds effectively reduce the number of BVTT nodes, thereby decreasing memory usage and computational costs. For complex scenes, e.g. Benchmark Intersection, not employing the enhanced bounds may result in an exceedingly high number of BVTT nodes, leading to the algorithm's crash. 

\DEL{
We have also developed a `na\"ive' GPU solution, serving as an ablation study, to validate the advantages of the techniques we proposed such as f-12 BVH and adaptive BVTT expansion. For more specific details, please refer to the supplementary material.
}

\subsection{Running Time Ratios}
Figure~\ref{fig:runtime-ratios} meticulously dissects the running time ratios across various computation stages for Benchmark Tools (left) and Benchmark Comet-Tool (right) on the NVIDIA GeForce RTX 4090. The results reveal that CPU/GPU Sync stands out as the most computationally intensive segment, constituting a substantial portion ($40\%-66\%$) of the total running time. This aligns with the iterative nature of our algorithm, which necessitates frequent data transfers from the GPU to the CPU. These transfers are crucial for calculating BVTT front expansion depths and determining iteration termination criteria.
\HL{
Typically, CUDA kernel calls operate asynchronously. However, before launching the CUDA kernel for BVTT expansion, the CPU must specify the number of threads, It is a value closely tied to the number of BVTT nodes, and needs to be transmitted from the GPU to the CPU. This operation disrupted the asynchronous execution between CPU and GPU, incurring additional synchronization costs.
}

\begin{figure}[t]
	\centering
	\includegraphics[width=0.82\linewidth]{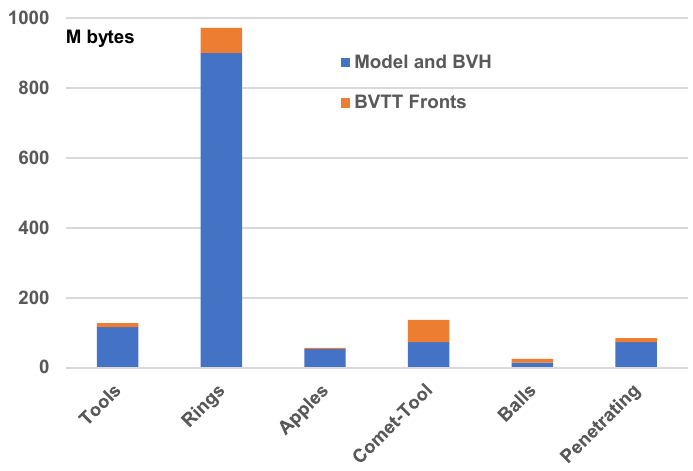}
    \caption{{\bf Memory Overhead:} The figure compares the GPU memory overhead of our algorithm on different benchmarks. The memory overhead can be broken down into two parts: a static part (for Model and BVH) and a dynamic part (for BVTT fronts). 
	}
	\label{fig:memory-ratios}
\end{figure}

\subsection{Memory Overhead}
Figure~\ref{fig:memory-ratios} presents a comparative overview of the GPU memory overhead for different benchmarks. The memory overhead is dissected into two components: a static part (for Model and BVH) and a dynamic part (for BVTT fronts). The dynamic part showcases the peak memory utilization for each benchmark. 
The first part involves the BVH tree nodes and information about the model, which is stored on the leaf nodes of the BVH. This part is essential for almost all BVH-based distance query algorithms. The second part pertains to the additional memory consumption introduced by our algorithm, which we want to highlight is very low. 
 
Even in the most intricate scenario (Benchmark Rings with $15$M triangles), the BVTT fronts consume less than $72$M GPU memory.
\HL{We utilized around $900$M ($500$M+$400$M) to store the BVH tree (leaf+non-leaf nodes), and $71.6$M for BVTT fronts.}
This starkly contrasts with earlier GPU algorithms~\cite{Tang18,BVHCD18}, which often incurred approximately $500$M memory usage \HL{for BVTT fronts}. Comparable memory reduction is observed across various benchmarks, attesting to the efficiency of our algorithm in managing GPU memory resources.

\DEL{
The inherent temporal coherency of deformable bodies is strategically exploited using a lightweight approach. Specifically, we record model vertices ($V_a$, $V_b$) in proximity to the nearest point pair ($P_a$, $P_b$) from the previous frame. In the subsequent frame, these recorded vertices ($V_a$, $V_b$) aid in initializing estimates for maximum and minimum distances, using $Dist(V_a, V_b)$. This strategy incurs minimal computational cost and yields notable benefits. Notably, this temporal coherency utilization extends beyond deformable bodies, proving advantageous for rigid bodies undergoing continuous motion. Empirical results demonstrate a 
$10\%-12\%$ speedup compared to an implementation devoid of this temporal coherency strategy. 
}

Additional details about certain algorithms and performance data for benchmarks can be found in the supplementary materials.

\section{Conclusion and Limitations}
\label{sec:conclude}

We introduce gDist, a GPU-based method for distance computation, leveraging an iterative algorithm pipeline (Algorithm~\ref{alg:overall}), tight distance bounds for AABBs (Section~\ref{sec:AABB}), parallel BVTT expansion (Algorithm~\ref{alg:BVTT}), adaptive expansion depth computing (Section~\ref{sec:depth}), and other GPU optimization techniques (Section~\ref{sec:parallel}). The efficacy of gDist is demonstrated through extensive benchmarking on complex datasets comprising $19.4K-15M$ triangles. Our results showcase remarkable speedups \DEL{(ranging from $1$ to $3$ orders of magnitude)} and substantially reduced memory overhead compared to previous algorithms. 
This work introduces a novel algorithm explicitly tailored for GPU parallel mesh-to-mesh distance queries. To the best of our knowledge, no publically available GPU algorithms provided satisfactory performance for this specific task before. Therefore, our approach makes a valuable contribution by addressing \HHHL{the challenge}.

Our approach has several limitations. While gDist represents a significant advancement, certain limitations warrant consideration. First, the acceleration achieved by our algorithm is contingent on scene complexity. In simpler scenes with fewer triangles, the potential effectiveness of our parallel acceleration might be constrained. Additionally, the culling effectiveness of BVTT depends on the scenario. For instance, in models like two large concentric spheres where numerous triangles share identical maximum/minimum distances, BVTT culling might not eliminate many detection branches, thereby limiting our acceleration performance. This limitation is inherent, and even CPU algorithms struggle in similar scenarios. Another constraint arises in very complex scenes where GPU memory usage surpasses the capacity of commodity GPUs. In such cases, performing distance calculations in batches becomes necessary, leading to a noticeable decline in performance.

\HHL{In paper~\cite{Quinlan94}, relative error is used to improve computational efficiency, leading to conservative estimates of minimum and maximum distances. Although this method may enhance culling and computation speed in some scenarios, our primary focus in this paper is the precise distance computation between objects. To ensure accuracy, we chose not to use relative error estimation. We will consider integrating both methods in future work to improve efficiency while maintaining accuracy.}

Several promising avenues for future research emerge. Overcoming the identified limitations is a priority, and exploring the computational potential of multiple GPUs through judicious data/task partitioning for parallel distance computation in large-scale scenes presents an intriguing prospect~\cite{pcloth20}. \DEL{This multi-GPU approach holds the potential to enhance both scalability and performance, opening up new possibilities for advancing distance computation methodologies.} \HHL{Other intriguing avenues for future research involve extending our current 1-to-1 distance query framework to accommodate 1-to-N distance queries, exploring distance computations between SDF-based models~\cite{Macklin2020}, as well as using relative errors~\cite{Quinlan94} for faster and conservative distance computing. }

\begin{acks}
This work was supported by the Leading Goose R\& D Program of Zhejiang under Grant No. 2024C01103.
\end{acks}


\bibliographystyle{ACM-Reference-Format}
\bibliography{main}
\clearpage

\appendix

\begin{figure}[h]
	\centering
	\includegraphics[width=0.88\linewidth]{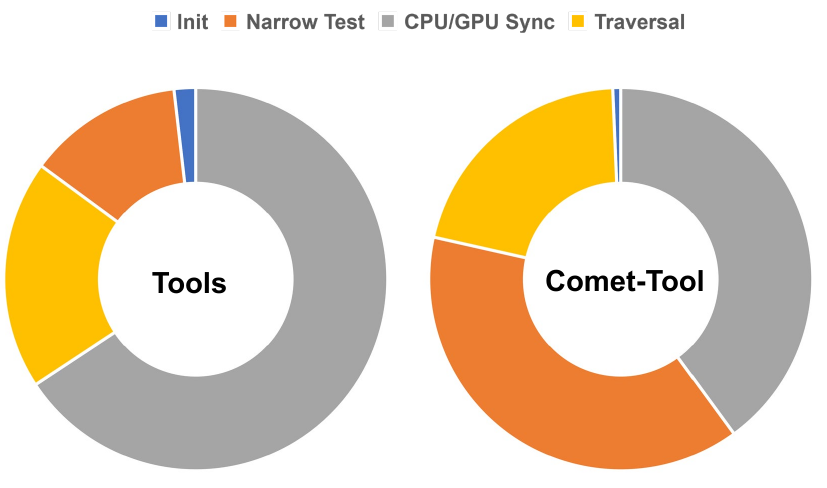}
    \caption{{\bf Runtime Ratios:} The figure shows the running time ratios of different computing stages: data initialization, BVTT traversal, narrow phrase testing, and CPU/GPU sync. These data are collected by running our system for Benchmark Tools (left) and Benchmark Comet-Tool (right).}
	\label{fig:runtime-ratios}
\end{figure}

\begin{figure}[h]
	\centering
	\includegraphics[width=1\linewidth]{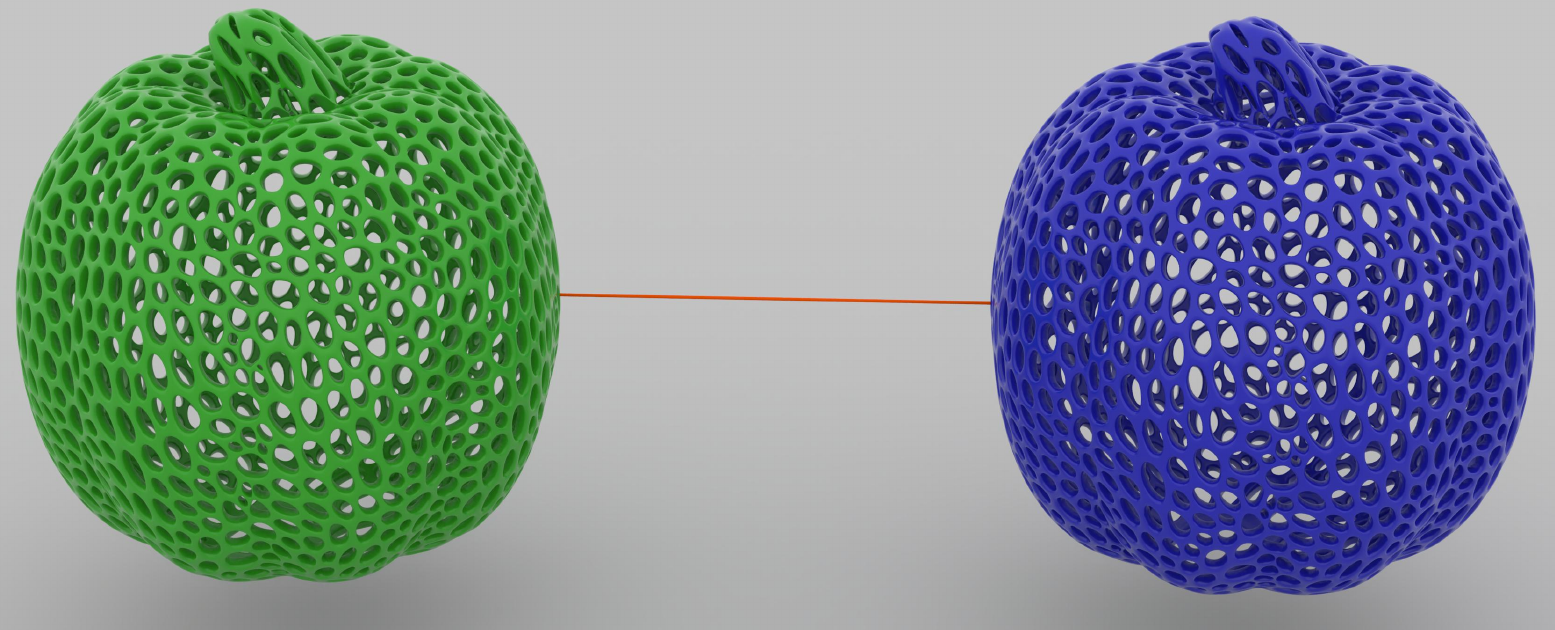}
    \caption{{\bf Benchmark Apples:} 
    Two Voronoi apples are moving apart from each other. Each apple comprises $720K$ triangles.
    \HHHL{Compared to the `na\"ive' GPU implementation, we observe about $12$X speedups} along the moving trajectory on an NVIDIA GeForce 4090.
	}
	\label{fig:bench-3}
\end{figure}

\begin{figure}[h]
	\centering
	\includegraphics[width=0.69\linewidth]{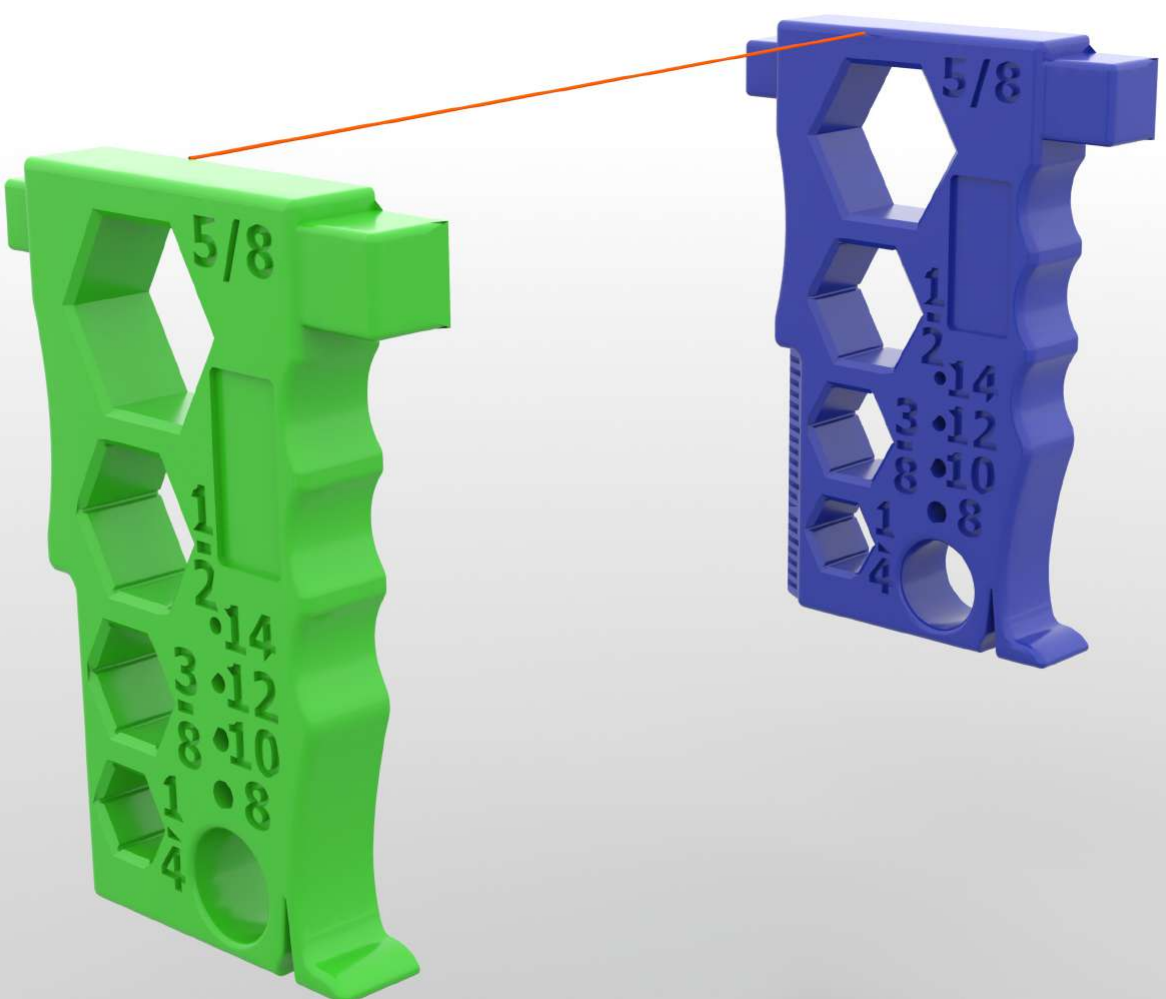}
    \caption{{\bf Benchmark Tools:} This benchmark features a tool moving in parallel to another tool. Each tool consists of $1M$ triangles. \HHHL{Compared to the `na\"ive' GPU implementation,  we observe about $8.5$X speedups} on an NVIDIA GeForce 4090.
        }
	\label{fig:bench-1}
\end{figure}

\begin{figure}[H]
	\centering
	\includegraphics[width=0.58\linewidth]{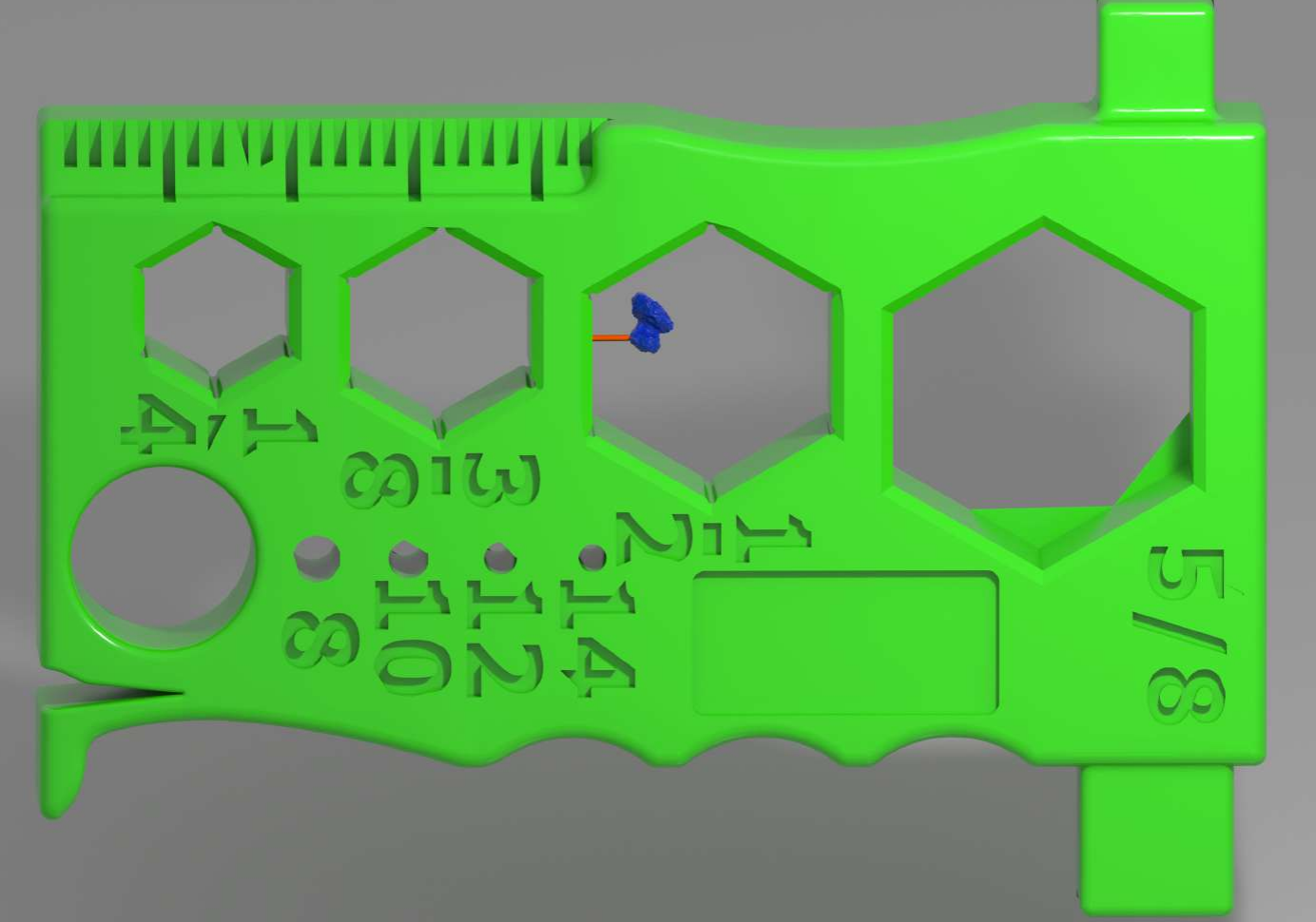}
    \caption{{\bf Benchmark Comet-Tool:} This benchmark involves a comet model, comprising $1$M triangles, moving through a hole in a tools model, also with $1$M triangles. The two objects are at quite different scales. \HHHL{Compared to the `na\"ive' GPU implementation, we observe above $440$X speedups} on an NVIDIA GeForce 4090.
	}
	\label{fig:bench-5}
\end{figure}

\begin{figure}[H]
	\centering
	\includegraphics[width=0.55\linewidth]{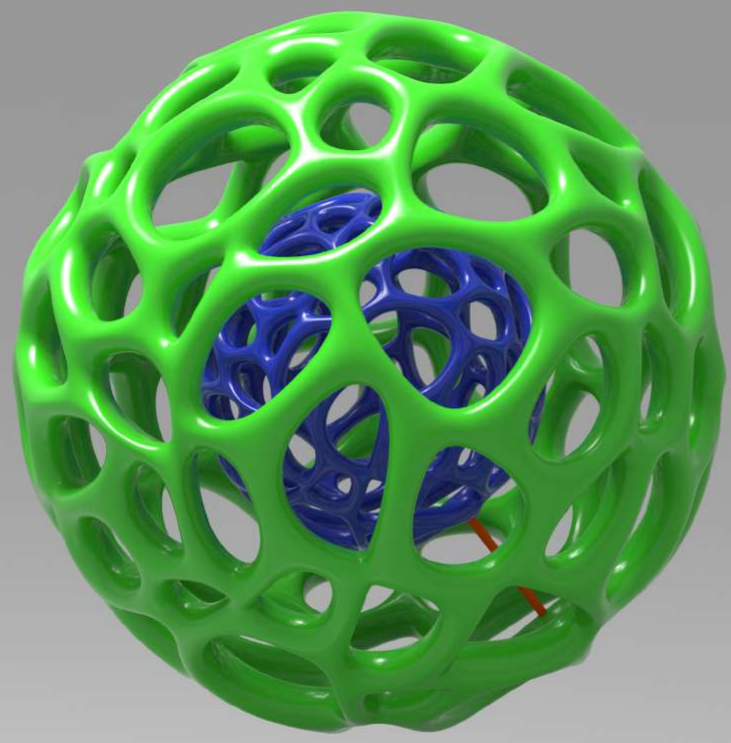}
    \caption{{\bf Benchmark Balls:} A smaller Voronoi ball is enclosed by a larger Voronoi ball, and both are rotating synchronously about the same axis. Each apple comprises $130$K triangles. \HHHL{Compared to the `na\"ive' GPU implementation, we observe about $9.3$X speedups} on an NVIDIA GeForce 4090.
	}
	\label{fig:bench-6}
\end{figure}

\begin{figure}[H]
	\centering
	\includegraphics[width=0.45\linewidth]{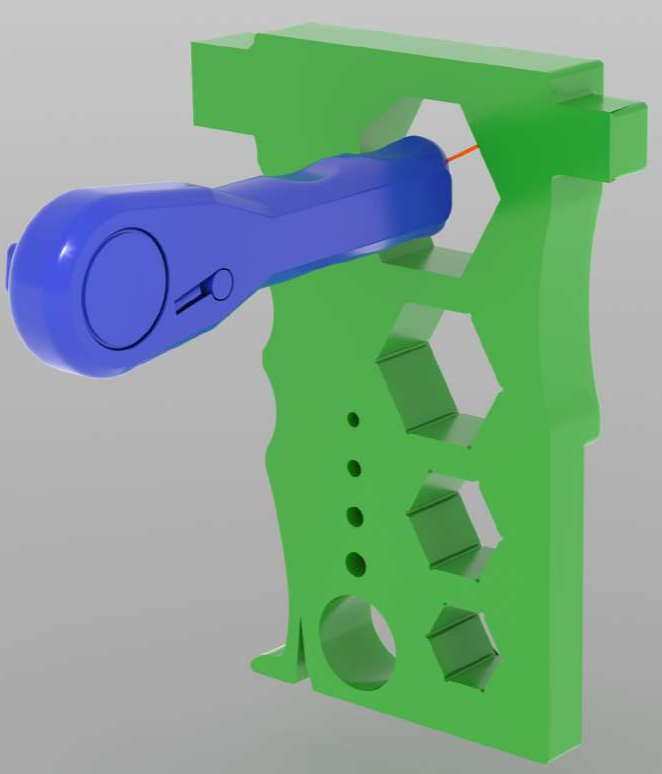}
    \caption{{\bf Benchmark Penetration:} A \HL{ratchet} (with $1$M triangles) moves through a hole in a base tool (also with $1$M triangles). \HHHL{Compared to the `na\"ive' GPU implementation, we observe above $377$X speedups} on an NVIDIA GeForce 4090.
	}
	\label{fig:bench-7}
\end{figure}

\begin{figure}[H]
	\centering
	\includegraphics[width=0.8\linewidth]{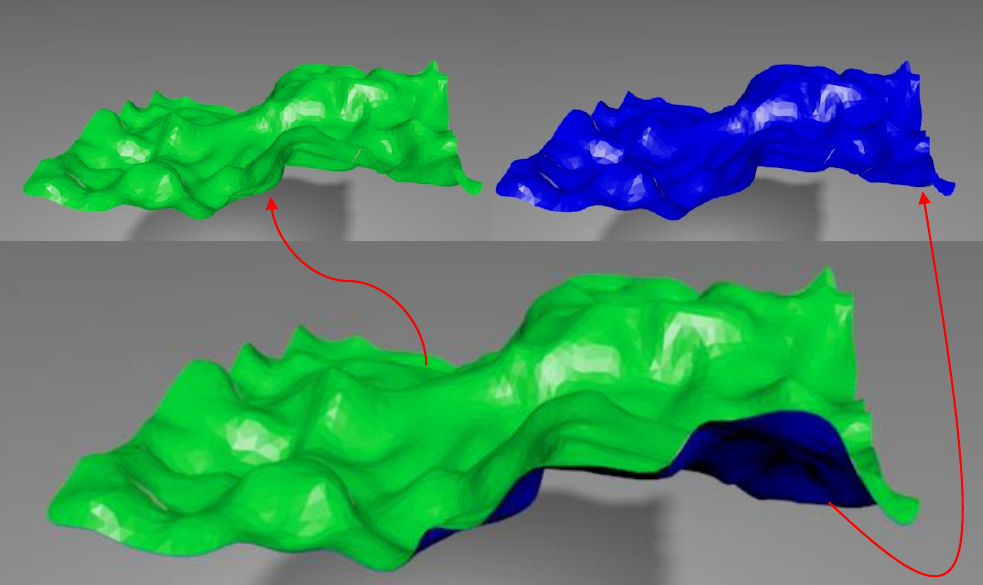}
    \caption{\HL{{\bf Benchmark Terrains:} Two terrain meshes (each with 9.7K triangles) are in close proximity to one other. \HHHL{Compared to the `na\"ive' GPU implementation, we observed about $13.5X$ speedups} on an NVIDIA GeForce 4090. }
	}
	\label{fig:bench-12}
\end{figure}

\begin{figure}[H]
	\centering
	\includegraphics[width=0.65\linewidth]{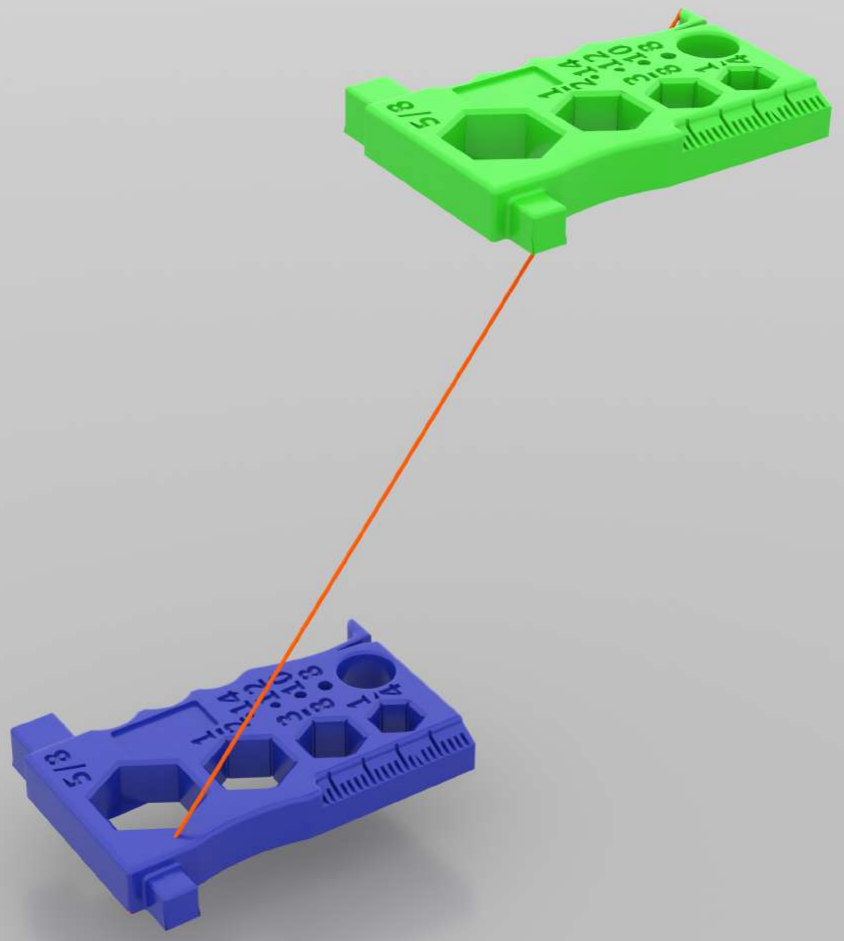}
    \caption{{\bf Benchmark MaxDist:} This benchmark mirrors the configuration of the Tools benchmark. However, in this case, the focus is on computing the maximum distance instead of the minimum distance. The average query time is about $0.25$ milliseconds per frame on an NVIDIA GeForce 4090.
	}
	\label{fig:bench-8}
\end{figure}

\begin{figure}[H]
	\centering
	\includegraphics[width=0.85\linewidth]{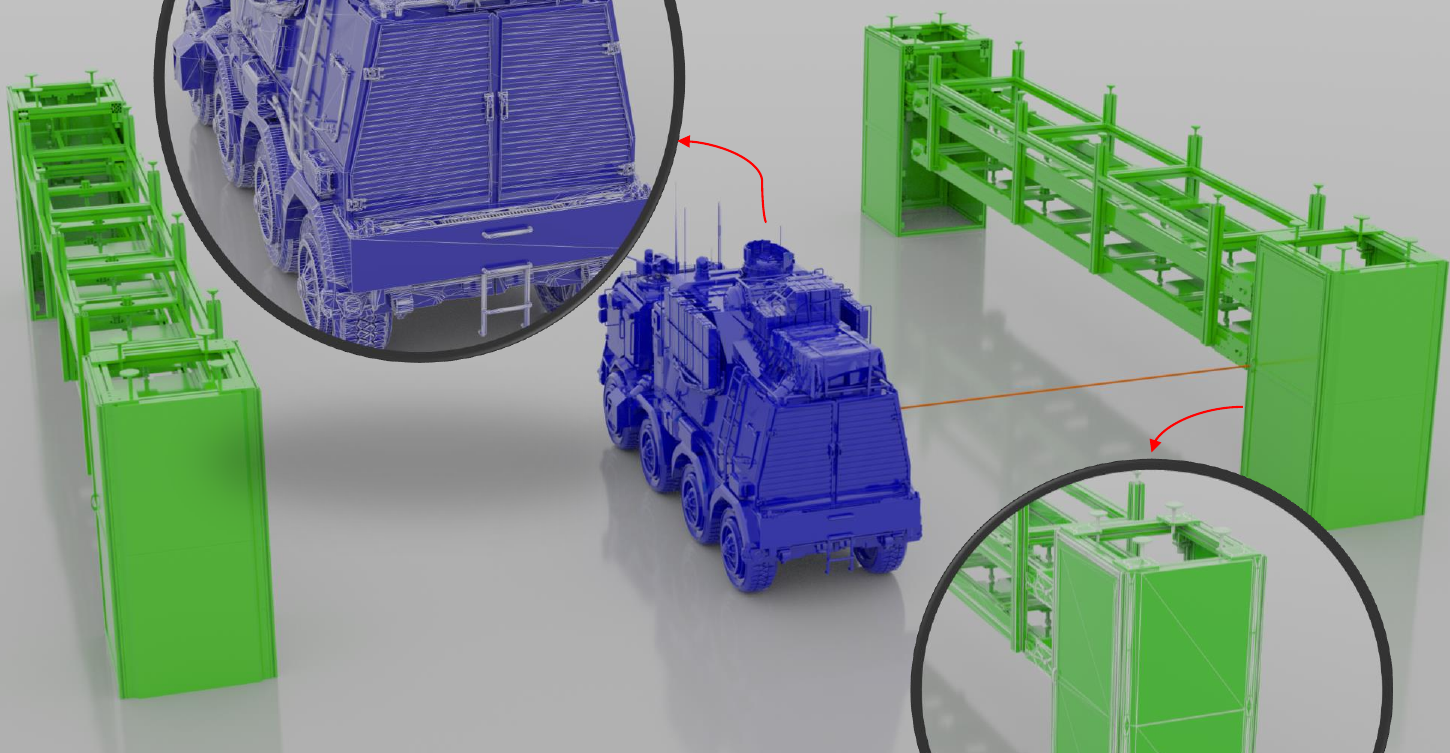}
    \caption{\HL{{\bf Benchmark Truck:} 
     A truck ($1.1$M triangles) is traveling between two rows of shelves  ($2.41$M triangles), and the objects in this scenario display non-uniform tessellation. \HHHL{Compared to the `na\"ive' GPU implementation, we observed about $835X$ speedups} on an NVIDIA GeForce 4090.}
	}
	\label{fig:bench-11}
\end{figure}

\begin{figure}[H]
	\centering
	\includegraphics[width=1.0\linewidth]{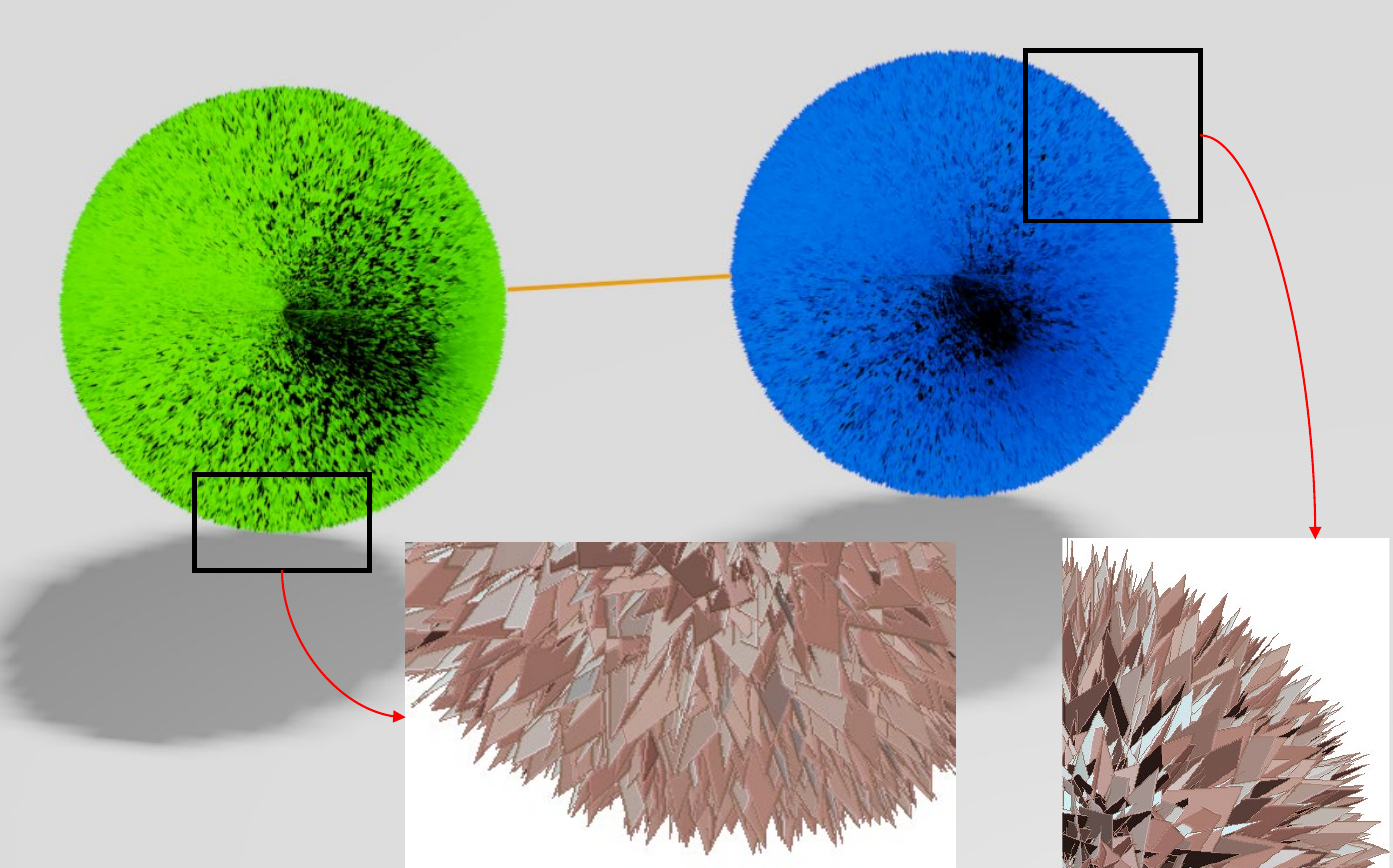}
    \caption{{\bf Benchmark Intersection:} 
    Each model comprises $1$ million triangles and intersects at a pre-specified point. 
    \HHHL{The `na\"ive' GPU implementation requires approximately $3778$ seconds for minimum distance computation between the two models, while our GPU algorithm accomplishes the same task in a mere $0.59$ milliseconds on an NVIDIA GeForce 4090, which is about $6,403$X faster.}
	}
	\label{fig:bench-10}
\end{figure}

\begin{figure}[H]
	\centering
	\includegraphics[width=1\linewidth]{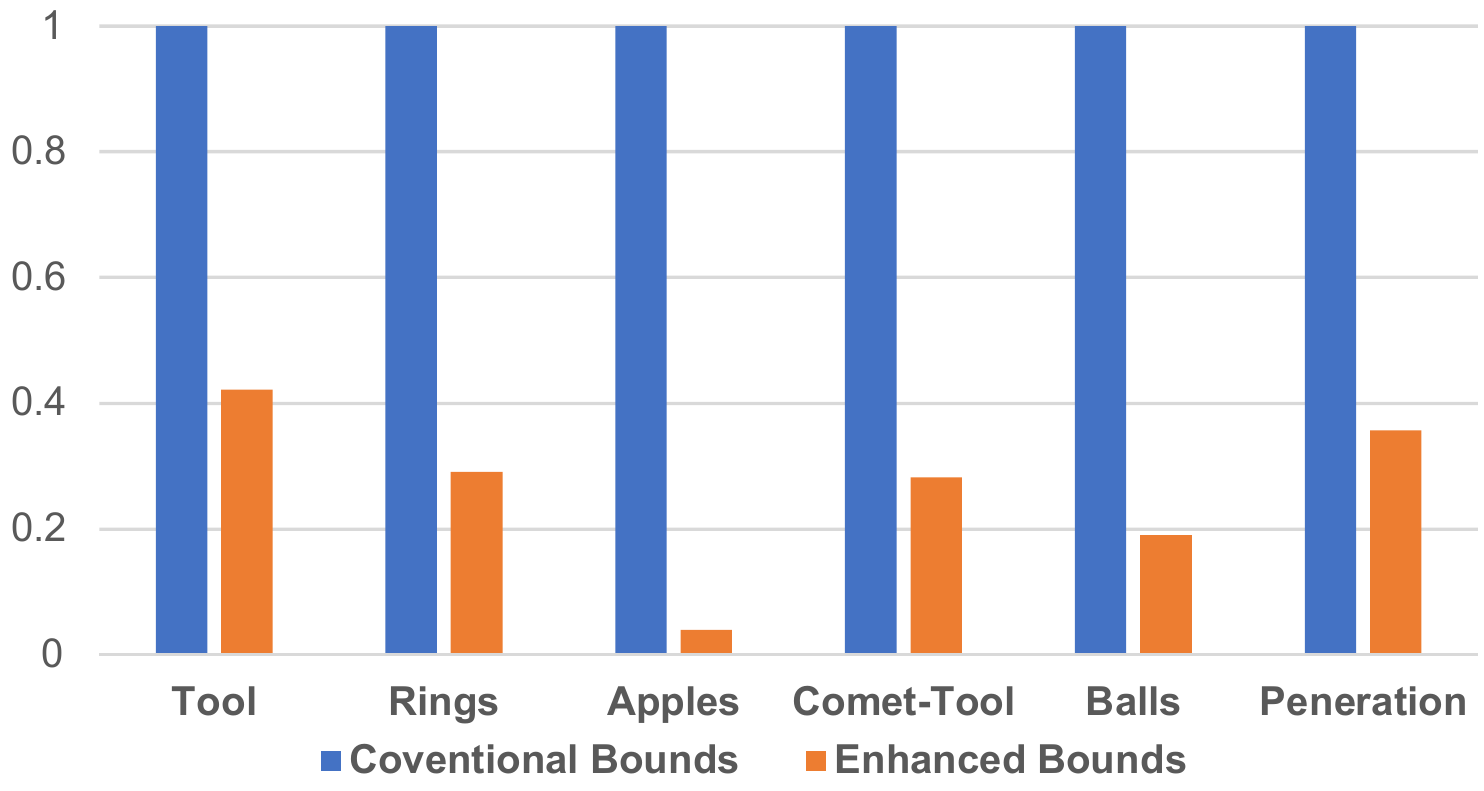}
    \caption{{\bf Ablation Study 2:} This figure presents another ablation study: Through the use of conventional distance bounds and our enhanced distance bounds, we can observe the changes in the number of BVTT nodes.
    We set the number of nodes when using the conventional bounds as $1$, and the number when using the enhanced bounds as its proportion. The culling rates in all scenarios are less than $45\%$. In Benchmark Apples, the number can be less than $4\%$. This experiment clearly demonstrates how the enhanced bounds effectively reduce the number of BVTT nodes, thereby decreasing memory usage and computational costs.
    }
	\label{fig:ablation2}
\end{figure}
\end{document}